# A wash-free, dip-type fiber optic plasmonic (DiP) assay for sub-zeptomole analyte detection

Bandaru Ramakrishna[1], Vani Janakiraman[2] and V V R Sai[1,*]

[1]Biomedical Engineering Laboratory, Department of Applied Mechanics, and [2]Department of Biotechnology, Indian Institute of Technology (IIT) Madras, Chennai, India - 600036.
* Corresponding author: vvrsai@iitm.ac.in

**ABSTRACT:** This study demonstrates a rapid, wash-free dip-type fiber optic plasmonic sandwich (DiP) assay capable of zeptomole analyte detection and 6-orders of wide dynamic range. The DiP assay is realized on a compact U-bent fiber optic probe surface by dipping the antibody functionalized probe into a mixture of sample solution and AuNP labeled reagent. U-bent fiberoptic probes with a high evanescent wave absorbance sensitivity allow detection of the high extinction gold nanoparticle (AuNP) labels in terms of a drop in the light intensity, which is measured with the help of a pair of LED and photodetector (PD). This simple and low-cost DiP assay gave rise to unprecedented detection limit down to 0.17 zeptomole of human immunoglobulin G (HIgG) in 25 μl buffer solution within an assay time of 25 min. Further, silver enhancement of AuNP labels over 5 min resulted in a limit of quantitation (LoQ) down to 0.17 zeptomole (~100 molecules in 25 μl). The DiP assay with high sensitivity and low detection limits within an assay time of 30 min demonstrated here could facilitate low cost point-of-care diagnostics as well as high-throughput systems.

Clinical diagnosis, at present, significantly utilizes conventional diagnostic procedures such as enzyme linked immunosorbent assay (ELISA)[1,2] and chemiluminescence immunoassays (CLIA)[3] due to their relatively high sensitivity, specificity and reliable detection of target biomarkers. However, these techniques also involve laborious and time-consuming experimental procedures, executed by either high-end instrumentation or skilled professionals. In addition, the challenges in realizing ultra-low analyte detection limits are some of the limitations as discussed elsewhere[4,5,6]. In recent years, several label-free optical diagnostic technologies based on surface plasmon resonance (SPR)[7], bio-layer interferometry (BLI)[8] or resonant waveguide grating (RWG)[9] have been developed for high throughput chemical and bimolecular analysis, which involve only minimal processing steps. The success of these commercialized technologies has been mostly limited to research laboratories mainly due to their high installation and running costs. Alternately, numerous nanotechnology based labeled diagnostic assays, such as plasmonic ELISA[12], bio-barcode assay[13] and Simoa[14] technologies capable of femto or attomolar analyte sensitivities have emerged for commercial applications. However, similar to the label-free technologies, these technologies also suffer from higher installation and/or running costs for wide-scale deployment, especially in resource poor environment.

On the other hand, lateral flow assays (LFA)[15], which exploit gold nanoparticles (AuNP) labels in a sandwich assay format, remain as the highly successful point-of-care (PoC) diagnostic technology. This is mainly due to their exceptional features including low-cost (less than a dollar), rapid analysis and naked eye based detection of biomarkers. However, there are several constraints including low sensitivity, inability to quantify specific and nonspecific binding and requirement of a separate reader instrumentation that limit its scope mostly to qualitative analysis[16]. Despite exploiting the catalytic properties of AuNP labels, only an order of improvement in detection limits were reported[17]. As per the ASSURED criterion recommended by World Health Organization (WHO)[18], ideal PoC devices should not only inherit the simplicity of LFA such as all-in-one, one-step and wash-free assay, but also overcome the above constraints. In addition, the sample volumes as low as 25 μl (collected by finger prick method) and femto or attomolar analyte detection sensitivities could be highly desirable for clinical applications including early diagnosis of certain diseases. One approach to develop such a PoC device is to design a one-step biosensor by realizing a plasmonic assay (as in LFA) on an efficient optical transducer. Evanescent wave based fiber optic transducers have been shown to be highly promising for realization of plasmonic biosensors with femtomolar (fM) analyte sensitivity. The high extinction coefficient of AuNP labels and efficient light-matter interactions at the fiber sensor surface could potentially bring down the analyte detection limits to unprecedented attomolar or zepto mole regime.

In this direction, we propose an optical biosensor strategy that exploits the high extinction coefficient of AuNP labels in a plasmonic sandwich assay for analyte detection by means of a highly sensitive optical absorbance sensor in the form of a U-bent fiber optic probe. Figure 1A presents a schematic for the working principle of a dip-type U-bent fiberoptic plasmonic absorbance (DiP) assay. A compact U-bent probe is fabricated from a fused silica fiber (200 μm core diameter) by decladding, and bending under a butane flame to obtain a bend diameter of 1.5 mm (see Supporting Information). A U-bent probe efficiently converts the large number of lower order modes concentrated around the optical axis into higher order modes, thus allowing a large fraction of light available at the core surface. This leads to an improved evanescent wave (EW) based light-matter interaction at the core-medium interface at the bent region. This results in probes with significantly higher absorbance sensitivity[20,21] to detect a small number of surface-bound AuNP. Presence of AuNP on the fiber core surf-

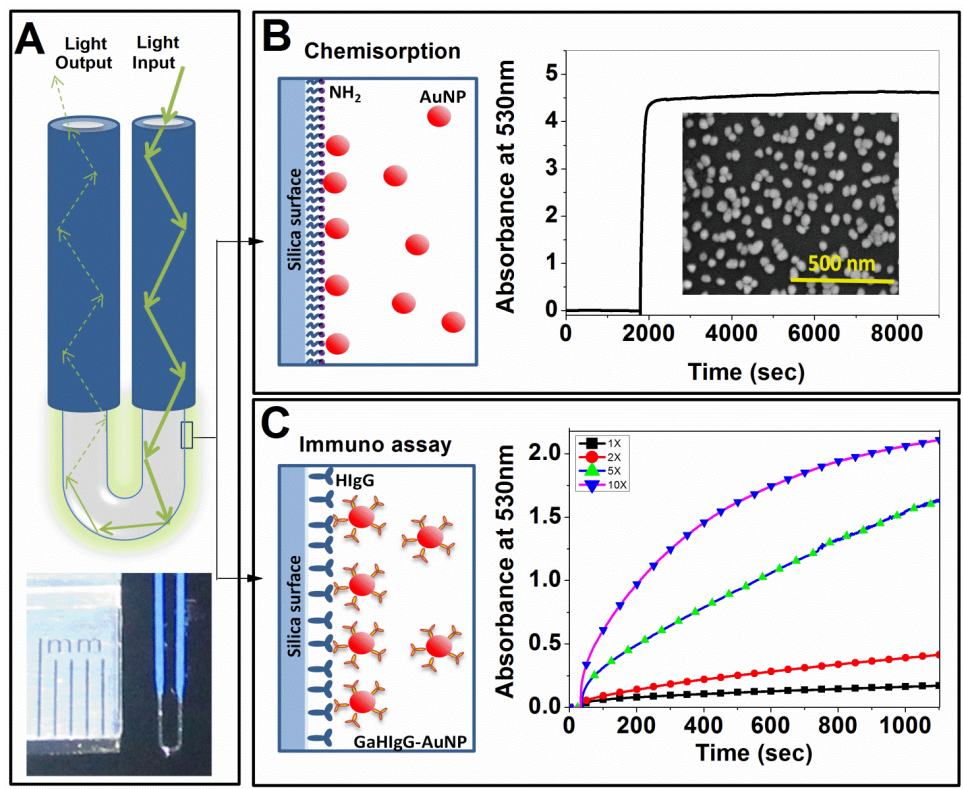

**Figure 1: Ultrasensitive detection of AuNP labels using dip-type U-bent fiberoptic sensor probes**. **A.** Schematic showing the working principle of DiP assay and a photographic image of a U-bent fiberoptic probe. The optical power loss in the U-bent region due to EW based absorbance by the AuNP bound to the probe surface is depicted in terms of thickness of the light ray. **B.** Saturated EW based absorbance response due to chemisorption of AuNP to amino-silanized U-bent fiber probe surface. Inset: The scanning electron microscopic images showing a dense surface coverage of AuNP. **C.** Absorbance response obtained from HIgG antibody functionalized U-bent probe as a result of direct assay when subjected to 1× to 10× concentrations of complementary GaHIgG antibody conjugated AuNP (GaHIgG-AuNP) to evaluate the concentration effects.

-ace in the bend region gives rise to a significant optical power loss at the other end of the U-bent of the probe (Fig 1A) upon its EW based light interaction with the AuNP. The unbound AuNP floating in the medium do not elicit any response due to the fact that the EWs are a near-field phenomenon with its depth of penetration limited to the order of the wavelength of the light.

To demonstrate the efficiency of U-bent probes to detect the presence of AuNP labels, amine functionalized U-bent probes were exposed to AuNP solution (see Figure S1, Supporting information) and the drop in the intensity of light at the photodetector (PD) end was monitored. Figure 1B shows the real-time absorbance response due to formation of a chemisorbed monolayer of AuNP measured using a narrow band green light source (laser at 532 nm) and a photodetector (PM100, Thorlabs Inc. USA).. By leveraging the high absorbance sensitivity of the probes and the large dynamic range and high resolution of the photodetector, a drop in light intensity as high as more than four orders was observed for a saturated surface coverage (~150 particle/μm$^2$) of AuNP, while AuNP surface densities as low as 6 - 7 particles/μm$^2$ lead to a significant rise in absorbance of 0.1 units (Figure S2, Supporting information).Thus, the results show that the absorbance response from the probe varies proportionally to the surface density of the AuNP labels. Subsequently, a direct immunoassay was realized to obtain a deeper understanding of the influence of the concentration of AuNP labels on assay times and the magnitude of the sensor response (Figure 1C, and Figure S3, Supporting information).U-bent fiber probes were functionalized witn HIgG and subjected to goat anti-human IgG (GaHIgG, $F_c$ specific) antibody conjugated AuNP labels (AuNP reagent), The probes were subjected to different concentrations of AuNP reagent from 1× to 10× (bare AuNP OD$_{530nm}$ = 0.75 arb. units). A saturated absorbance response of ~2 units within 20 min was obtained with 10× AuNP reagent, while 5×, 2× and 1× reagents show only ~77%, 19% and 8% of that from 10× concentration respectively. These results demonstrate a high sensitivity of the U-bent probes for an immunoassay with plasmonic AuNP labels.

Subsequently, a two-step sandwich assay was realized on a dip-type U-bent probe as a proof-of-the-concept for DiP assay (Figure 2). As the first step, HIgG target analyte was captured onto $F_c$ specific GaHIgG conjugated AuNP labels, mainly to reduce assay times by exploiting liquid phase homogeneous mixing (Figure 2A). AuNP reagent (10×, 25 μl) containing a total number of ~7.5×10$^{10}$ AuNP (ca. 50 antibodies per AuNP)[22] was added to a sample volume of 25 μl containing 10 pg/ml (70 fM) of HIgG analyte (10$^6$ molecules) in a 0.2 ml microcentrifuge tube. A high concentration of AuNP reagent nullifies any hook effect[23] and ensures at least one AuNP for each analyte molecule to form HIgG-GaHIgG-AuNP immunocomplex for their detection. In the second step, a U-bent fiberoptic probe functionalised with $F_{ab}$ specific GaHIgG was dipped into the microcentrifuge tube to allow the formation of sandwich immunocomplexes on a fiber probe surface. The absorbance of light due to the presence of AuNP labels was monitored using a pair of narrowband green LED and a PD (Figure 2A and Figure S3, Supporting information). Figure 2C shows temporal response obtained from a probe subjected to an analyte concentration of 10 pg/ml over 20 min.

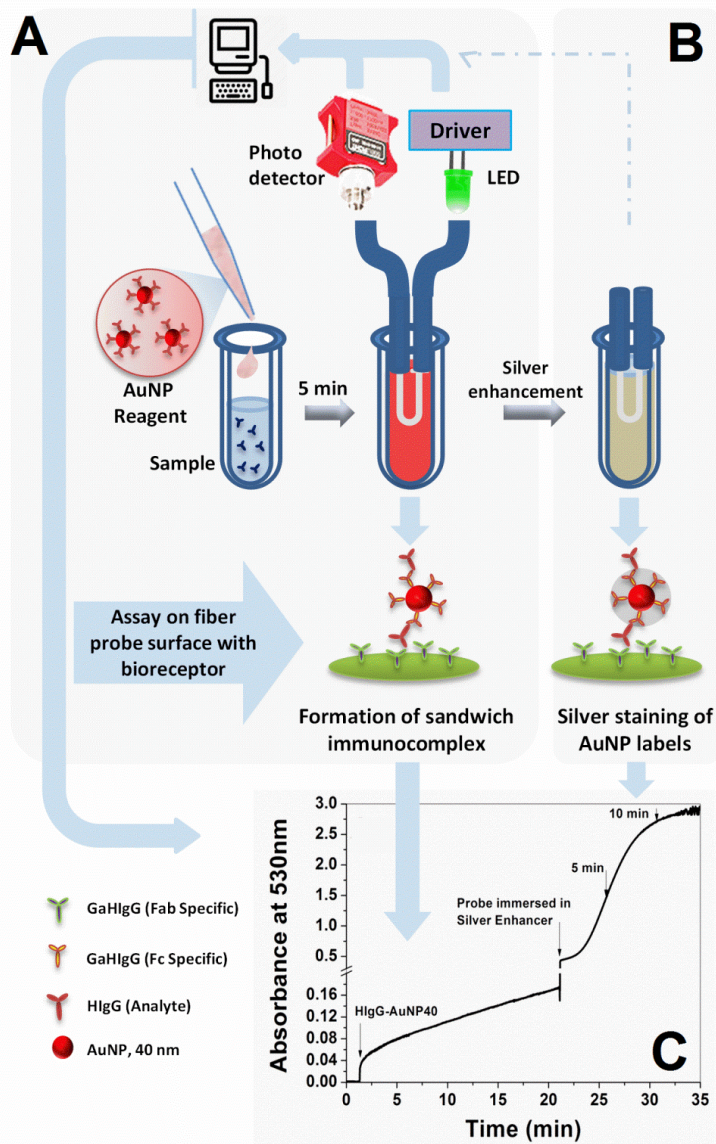

**Figure 2: Schematic showing analyte detection using DiP assay. A.** Step 1: A sample solution (25 µl) in a 0.2 ml microcentrifuge tube containing HIgG in phosphate buffer saline (PBS) is mixed with a labelled reagent (25 µl) containing AuNP conjugated with GaHIgG ($F_c$ specific) antibodies to allow bioaffinity reaction for 5 min. Step 2: A U-bent fiberoptic probe coupled between a pair of LED (LED528EHP, 7 mW, Thorlabs Inc. USA) and photodetector (10 pW-50 mW, S150C, Thorlabs Inc., USA) is dipped in the sample-reagent solution. The fiber core is biofunctionalized with GaHIgG ($F_{ab}$ specific) antibodies to allow formation of plasmonic sandwich immunocomplex on the probe surface for 20 min. (Note: The distal ends of the fiber probe are connected to LED and PD through bare fiber adaptors and SMA connectors, to avoid bulky optomechanical accessories). **B.** The fiber probe is dipped in silver enhancer solution to amplify the plasmonic absorbance signal by catalytic reduction of silver around AuNP. **C.** The sensor response depicted in terms of absorbance by normalizing the optical intensity signal from PD with respect to its initial value at the beginning of the experiment.

As the third step, silver enhancement over AuNP labels by catalytic reduction was adopted to amplify the plasmonic absorbance response[24]. Probes were silver stained by dipping them in phosphate buffer saline (PBS) twice followed by a mixture of silver enhancer and precursor solution (Sigma Chemicals, USA) in 1:1 v/v ratio for 15 min (Figure 2B). Figure 2C shows an exponential rise in absorbance response for the first 5 min followed by saturation at the end of 15 min. Scanning electron microscope (SEM) images show a controlled growth of silver over AuNP forming islands as large as 5 times the size of AuNP in the initial 5 mins, followed by uncontrolled, non-specific silver deposition due to photo catalysis at later time points (Figure S4, Supporting information). Silver staining for 5 min consistently gave rise to a reliable amplification of the sensor response.

Further, the sensitivity, dynamic range and detection limits of DiP assay were evaluated. GaHIgG ($F_{ab}$ specific) functionalized U-bent probes (>25 Nos) having a similar sensitivity (details in methods) were exposed to analyte concentrations over the range of 1 fg/ml (7 aM) to 1 ng/ml (7 pM) of HIgG (containing a total of ~$10^2$ to ~$10^8$ molecules respectively) in addition to control (0 fg/ml). DiP assay shows a distinct response proportional to the increasing analyte concentrations (n = 3) at the end of 20 min (Figure 3A and Figure S5, Supporting information). Absorbance response from control probes without HIgG was 0.047 ± 0.005, which is considerably lower than 0.074 ± 0.009 obtained for 1 fg/ml. It is interesting to note that the response due to non-specific binding of PEGylated GaHIgG-AuNP (control) is less than 2.5% of the maximum possible absorbance response due to the specific binding observed in Figure 1C. The analytical dose response shows a linear increase in absorbance between 1 fg/ml and 1 pg/ml with a sensitivity of 0.0148 $A_{530nm}$/log (fg/ml) ($R^2$=0.98) (Figure 3C). The sensor was able to consistently detect analyte concentrations as low as 1 fg/ml (LoD, ~7 aM or 0.17 zeptomole in 25 μl). However, the LoQ based on the response that is 10σ above the mean blank (control) value corresponds to 100 fg/ml. Hence, DiP assay with plasmonic labels alone offers 4-orders of dynamic range starting from 100 fg/ml (0.7 fM) to 1 ng/ml (7 pM). An extrapolation of the response curve shows the ability of the sensor to respond to analyte concentrations much higher than 1 ng/ml. To confirm the sensitivity of the DiP assay, these results were compared with that obtained from standard enzyme linked immunosorbent assay (ELISA) for the analyte concentrations from 1 fg/ml to 10 μg/ml, Analyte concentrations down to 10 ng/ml were detected using ELISA (Figure S6, Supporting information).

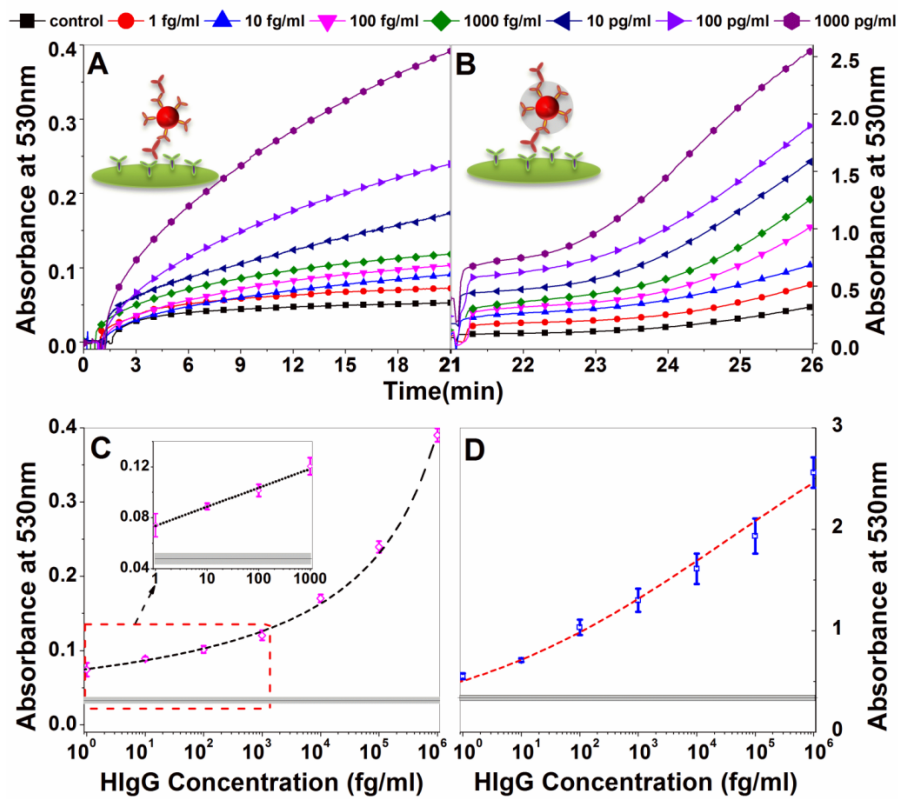

**Figure 3: Detection of ultralow analyte concentrations using DiP assay. A.** Temporal absorbance response obtained from U-bent probes due to the formation of plasmonic sandwich immunocomplex (GaHIgG-HIgG-GaHIgG-AuNP) on the fiber probe surface for HIgG concentrations from 1 fg/ml (7 aM) to 1 ng/ml (7 pM). **B.** Exponential rise in absorbance response when probes were dipped in silver enhancer solution for about 5 min. **C.** HIgG dose response curve obtained from a 20 min sandwich assay with 10× NP-GaHIgG labels over a dynamic range of 1 fg/ml to 1 ng/ml. Inset: Linear absorbance response (with $R^2 = 0.98$, n≥3) obtained from log concentrations of 1 fg/ml to 1 pg/ml of HIgG. **D.** Dose response curve for the corresponding fiber probes subjected to silver enhancement for 5 min.

Upon silver staining of AuNP labels on the fiber probes for 5 min, an exponential increase in the real-time absorbance response proportional to the analyte concentration was obtained (Figure 3B). While the control probe gave rise to an absorbance of 0.337 ± 0.017, a significantly higher response of 0.552 ± 0.03 was obtained for probes treated with 1 fg/ml analyte concentration. The dose response curve was considerably linear for the analyte concentrations under investigation with an absorbance sensitivity of ~0.281 $A_{530nm}$/log (fg/ml) ($R^2 = 0.91$), which is ~19 times higher than that of the sensitivity obtained with plasmonic labels alone (Figure 3D). The results from silver enhancement demonstrate a significant improvement in LoQ down to 1 fg/ml. For analyte concentrations higher than 1 pg/ml, a large standard deviation in the absorbance response was observed. This could be attributed to a wider distribution in the probable number of bound AuNP labels for higher analyte concentrations, which results in a wide variation in the absorbance response upon amplification by a large factor of ~20. Hence, silver enhancement could be most suitable to obtain dynamic range over the lower analyte concentrations from 1 fg/ml to 100 fg/ml, where the assay with AuNP labels alone lack enough sensitivity to quantify the analyte.

The results obtained in this study have multiple important attributes: (i) DiP assay allows rapid dip-type assay for detection of analytes (within 25-30 min) in a small sample volume under static (no-flow) conditions, hence offer compatibility with conventional micro centrifuge tubes and well-plates, (ii) LoDs down to 7 aM using low-cost optoelectronic instrumentation, (iii) A wide dynamic range between 7 aM and 7 pM with the help of either bare or silver stained AuNP labels. Further, the versatility of this technique lies in the ability to fine-tune the dynamic range for larger analyte concentrations by tailoring the extinction of plasmonic labels (e.g. AuNP <40 nm). A simple cost analysis estimates the fabrication cost of the fiberoptic probes at less than 1 US dollar ea., while the optoelectronic instrumentation and optical coupling units used in the proof-of-concept experiments cost less than 1000 USD (mainly S150C detector and PM100 console from Thorlabs Inc., which can be replaced with a low-cost PD). The reagents including antibodies and AuNP in an assay costs ~1 US dollar. Thus, the DiP assay offers ultrahigh sensitivity at ultralow cost of ~2 US dollar (see supporting information). Further optimization of assay protocols and adoption of an optimally designed microfluidic platform may hopefully lead to detection of ultra low concentrations. It is important to note that the DiP assay relies on accurate measurement of light intensity, which is influenced by the ambient conditions including temperature and refractive index of the sample-reagent solution. These effects can be compensated by employing a reference probe along with the test probe(s), which may possibly improve the detection limits as well. Owing to the above merits, the DiP assay is highly promising for realization of cost-effective and ultrasensitive PoC diagnostics as well as high-throughput analysis.

## ASSOCIATED CONTENT

**Supporting Information**

The Supporting Information is available free of charge on the ACS Publications website. Experimental methods, gold nanoparticles characteristics and conjugation with antibodies, SEM images of silver enhanced probes after sandwich assay, ELISA results and assay cost analysis are provided in here.

## AUTHOR INFORMATION

**Corresponding Author:**
+91 44 2257 4076, vvrsai@iitm.ac.in.

**Notes**
The authors declare no competing financial interests.

## ACKNOWLEDGMENT

BR thanks Ministry of Human Resource Development under Government of India, BIRAC-SRISTI, Indo-German Science and Technology (IGSTC) for the partial financial support. VVRS acknowledges the research grant from IIT Madras and IGSTC.

Table of Contents artwork:

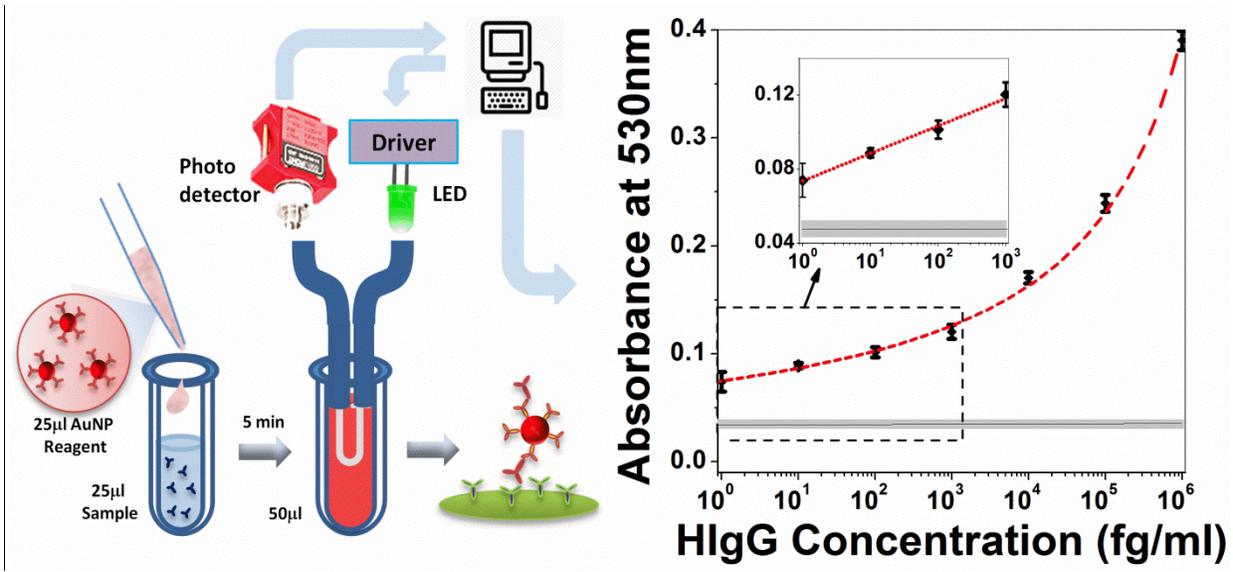